\documentclass[12pt,british]{article}
\usepackage[T1]{fontenc}
\usepackage[latin9]{inputenc}
\usepackage{geometry}
\usepackage{amssymb}
\usepackage{graphicx}

\makeatletter
\newcommand{\lyxrightaddress}[1]{
\par {\raggedleft \begin{tabular}{l}\ignorespaces
#1
\end{tabular}
\vspace{1.4em}
\par}
}

\def\vec#1{\mbox{\boldmath ${#1}$}}	
\def\d{\mathrm d}
\def\e{\mathrm e}
\date{}

\makeatother

\usepackage{babel}
\begin{document}

\title{\textbf{\Large Fundamental solutions of MHD Stokes flow}}

\author{J\={a}nis Priede}

\maketitle

\lyxrightaddress{{\small Applied Mathematics Research Centre, Coventry University,
Coventry, CV1 5FB, UK}}
\begin{abstract}
\noindent A simple analytical solution is obtained for the MHD stokeslet
in a homogeneous magnetic field. This solution represents the flow
past a small particle and can also be interpreted as the flow sufficiently
far away from a body of finite size. Fundamental solutions are found
in terms of velocity, pressure and scalar potential distributions
for the flows due to either a concentrated force or a current source.
The former consists of two basic solutions for the force parallel
and transverse to the magnetic field, respectively. All fundamental
solutions have the characteristic length scale of the Hartmann boundary
layer and two parabolic wakes developing along the magnetic field.
\end{abstract}

\section{Introduction}

The non-linear inertial effects in the flows of conducting liquids
may become negligible not only at low velocities but also in strong
enough magnetic fields \cite{BrTs70}. Numerical computation of such
flows is still complicated by the fine and intricate meshing required
to resolve the thin boundary and free shear layers that develop in
complex geometries. Owing to the linearity such flows can also be
computed using the boundary integral equation (BIE) techniques, which
requires only the surface but no volume meshing. This is approach
is well developed for Stokes flows in classical hydrodynamics \cite{Pozr92}
but not so in magnetohydrodynamics \cite{Tsi73a,Tsi73b}. The problem
is the so-called fundamental solution, which is essential for the
BIE formulation and describes the flow due to a concentrated point-force.
Such a fundamental solution, which plays the same role as the Coulomb's
law in the electrostatics and is known as a stokeslet in the hydrodynamics
\cite{HaSa80}, is still missing in the MHD. The aim of the present
study is to fill in this gap.

\section{Formulation of problem}

Consider an unbounded flow of a viscous electrically conducting liquid
in a homogeneous magnetic field $\vec{B}.$ The flow is due a body
which may be either dragged through the liquid by a constant force
$\vec{F}$ or discharging a direct current $I,$ which then interacts
with the externally applied magnetic field. The interaction of the
current with its own magnetic field is supposed to be negligible.
Sufficiently far away from the body, the flow is expected to be independent
of the body size and shape, and determined only by the total force
and the current. Thus, $\vec{F}$ and $I$ are assumed to be concentrated
at the body location $\vec{x}_{0},$ and the Dirac delta function
$\delta(\vec{x}-\vec{x}_{0})$ is used for their density distributions.
The origin of the coordinate system is set at the body location so
that $\vec{x}_{0}=\vec{0},$ and a simplified notation $\delta(\vec{x})\equiv\delta$
introduced. Neglecting the non-linear inertial forces, which supposes
either a creeping flow or a sufficiently strong magnetic field, the
flow is governed by the linearised Navier--Stokes equation with the
electromagnetic body term 
\begin{equation}
\eta\vec{\nabla}^{2}\vec{v}-\vec{\nabla}p+\vec{j}\times\vec{B}=-\vec{F}\delta,\label{eq:NS}
\end{equation}
where the velocity $\vec{v}$ is subject to the incompressibility
constraint $\vec{\nabla}\cdot\vec{v}=0.$ The current obeys both the
Ohm's law for a moving medium $\vec{j}=\sigma(-\vec{\nabla}\phi+\vec{v}\times\vec{B})$
and the charge conservation $\vec{\nabla}\cdot\vec{j}=I\delta.$ The
latter results in Poisson's equation for the electric potential $\phi:$
\begin{equation}
\vec{\nabla}^{2}\phi-\vec{B}\cdot\vec{\nabla}\times\vec{v}=-I\delta.\label{eq:phi}
\end{equation}
The resulting flow represents a superposition of mechanically and
electrically driven flows, which owing to the linearity of the problem
are mutually independent and, thus, are considered separately. Subsequently,
these two flow components are denoted by the indices $v$ and $\phi.$
Henceforth, we change to the dimensionless variables by using the
Hartmann layer thickness $d=B^{-1}\sqrt{\eta/\sigma}$ as the length
scale and $v_{0}=d^{2}F/\eta$ and $v_{0}=IB/\eta$ as characteristic
velocities for the flows driven mechanically and electrically. Other
quantities are scaled, respectively, by $p_{0}=v_{0}\eta/d,$ $j_{0}=v_{0}\eta/(d^{2}B),$
and $\phi_{0}=v_{0}dB,$ where $B=|\vec{B}|$ and $F=|\vec{F}|.$
The directions of the magnetic field and force are specified by the
unit vectors $\vec{\beta}=\vec{B}/B$ and $\vec{f}=\vec{F}/F.$ Governing
equations (\ref{eq:NS}) and (\ref{eq:phi}) can be written in dimensionless
form as
\begin{equation}
\vec{\nabla}^{2}\vec{v}-\vec{\nabla}p+\vec{\beta}\times(\vec{\nabla}\phi+\vec{\beta}\times\vec{v})=-\left[\begin{array}{l}
\delta\\
0
\end{array}\right]\vec{f};\qquad\vec{\nabla}^{2}\phi-\vec{\beta}\cdot\vec{\nabla}\times\vec{v}=-\left[\begin{array}{l}
0\\
\delta
\end{array}\right],\label{eq:vL}
\end{equation}
where the upper and lower cases on the r.h.s correspond to mechanically
and electrically driven flows. Applying the operators 
\begin{equation}
\vec{\mathcal{L}}_{v}^{v}=\vec{\nabla}\times\vec{\nabla}\times,\quad\mathcal{L}_{p}^{v}=\left[\vec{\nabla}^{2}\vec{\nabla}-\vec{\beta}\cdot\vec{\nabla}\vec{\beta}\right]\cdot,\quad\mathcal{L}_{\phi}^{v}=\vec{\nabla}\cdot\vec{\beta}\times\label{eq:Lv}
\end{equation}
to the first equation (\ref{eq:vL}) and taking into account the second
equation (\ref{eq:vL}), results in separate equations for each variable
\begin{equation}
\left[\vec{\nabla}^{4}-(\vec{\beta}\cdot\vec{\nabla})^{2}\right]\{\vec{v},p,\phi\}^{\{v,\phi\}}=\vec{\mathcal{L}}^{\{v,\phi\}}\delta,\label{eq:vpf}
\end{equation}
where the operator indices are expanded as follows 
\begin{equation}
\vec{\mathcal{L}}^{\{v,\phi\}}\equiv\left[\begin{array}{l}
\vec{\mathcal{L}}^{v}\vec{f}\\
\vec{\mathcal{L}}^{\phi}\mathcal{I}
\end{array}\right]\equiv\left[\begin{array}{l}
\{\vec{\mathcal{L}}_{v}^{v},\mathcal{L}_{p}^{v},\mathcal{L}_{\varphi}^{v}\}\vec{f}\\
\{\vec{\mathcal{L}}_{v}^{\phi},\mathcal{L}_{p}^{\phi},\mathcal{L}_{\phi}^{\phi}\}\mathcal{I}
\end{array}\right],\label{eq:L-rhs}
\end{equation}
 and the operators with the superscript $\varphi$ referring to the
electrically driven flow are 
\begin{equation}
\vec{\mathcal{L}}_{v}^{\phi}=\vec{\beta}\times\vec{\nabla},\quad\mathcal{L}_{p}^{\phi}=0,\quad\mathcal{L}_{\phi}^{\phi}=\mathcal{I}-\vec{\nabla}^{2}.\label{eq:Lf}
\end{equation}
Since both the l.h.s and r.h.s operators in (\ref{eq:vpf}) are linear
constant-coefficient operators and, thus, interchangeable with each
other, the solution of (\ref{eq:vpf}) can written as 
\begin{equation}
\{\vec{v},p,\phi\}^{\{v,\phi\}}=\vec{\mathcal{L}}^{\{v,\phi\}}G,\label{sol:vpf}
\end{equation}
 where $G$ is the fundamental solution satisfying 
\begin{equation}
\left[\vec{\nabla}^{4}-(\vec{\beta}\cdot\vec{\nabla})^{2}\right]G=\delta.\label{eq:G}
\end{equation}

\section{Fundamental solution}

For the sake of simplicity we further assume the $z$-axis to be directed
along the magnetic field so that $\vec{\beta}=\vec{e}_{z}$ and $(\vec{\beta}\cdot\vec{\nabla})^{2}\equiv\partial_{z}^{2}.$
Firstly, it is important to note that (\ref{eq:G}) can be factorised
in two ways as 
\begin{eqnarray}
\left[\vec{\nabla}^{2}\pm\partial_{z}\right]G\, & = & -G_{\pm},\label{eq:G-fct}\\
\left[\vec{\nabla}^{2}\mp\partial_{z}\right]G_{\pm}\!\!\! & = & -\delta.\label{eq:G-pm}
\end{eqnarray}
Secondly, the substitution $G_{\pm}=\e^{(\pm z-R)/2}G_{0},$ where
$R=|\vec{x}|$ is the spherical radius, reduces (\ref{eq:G-pm}) to
$\vec{\nabla}^{2}G_{0}=-\delta,$ whose fundamental solution is the
well-known Coulomb potential 
\begin{equation}
G_{0}(\vec{x})=1/4\pi R.\label{sol:G0}
\end{equation}
Adding and subtracting (\ref{eq:G-fct}) with different signs, results
in 
\begin{eqnarray}
\vec{\nabla}^{2}G & =-\frac{1}{2}(G_{+}+G_{-})= & -\cosh(z/2)\e^{-R/2}G_{0},\label{sol:d2G}\\
\partial_{z}G & =-\frac{1}{2}(G_{+}-G_{-})= & -\sinh(z/2)\e^{-R/2}G_{0}.\label{sol:dzG}
\end{eqnarray}
Equation (\ref{sol:dzG}) can be integrated as 
\begin{equation}
G(\vec{x})=\int\partial_{z}G\,\d z=-\frac{1}{8\pi}\left[E_{1}((R+z)/2)+E_{1}((R-z)/2)\right]+q(r),\label{sol:G}
\end{equation}
where $E_{1}(x)=\int_{x}^{\infty}t^{-1}\e^{-t}\,\d t$ is the exponential
integral and $q$ is a `constant' of integration. The latter is a
function of the cylindrical radius $r=\sqrt{R^{2}-z^{2}},$ and satisfies
the homogeneous counterpart of (\ref{eq:G}). It is determined as
$q(r)=-\ln r/4\pi$ by removing the logarithmic singularity $E_{1}(x)\sim-\ln x,$
which appears in (\ref{sol:G}) at the symmetry axis because $x=R-|z|\rightarrow0$
when $r\rightarrow0.$ The Laplacian and the $z$-component of the
gradient of $G$, which are required in (\ref{sol:vpf}), are given
by (\ref{sol:d2G}) and (\ref{sol:dzG}), respectively. The missing
$r$-component of the gradient can be obtained by differentiating
(\ref{sol:G}) and expressed in terms of the previous two quantities
and (\ref{sol:G0}) as
\begin{equation}
\partial_{r}G=-(R\vec{\nabla}^{2}G+z\partial_{z}G)/r-1/4\pi r.\label{sol:drG}
\end{equation}
For the following, it is important to note that the first term on
the r.h.s depends only on the cylindrical radius $r$ and decreases
inversely with $r$, whereas the second term similar to (\ref{sol:d2G})
and (\ref{sol:dzG}) depends also on the axial coordinate $z$ and
falls off exponentially at large $r.$

\subsection{Mechanically driven flow}

\paragraph{Axial force}

\begin{figure}
\begin{centering}
\includegraphics[bb=110bp 130bp 365bp 400bp,clip,width=0.33\columnwidth]{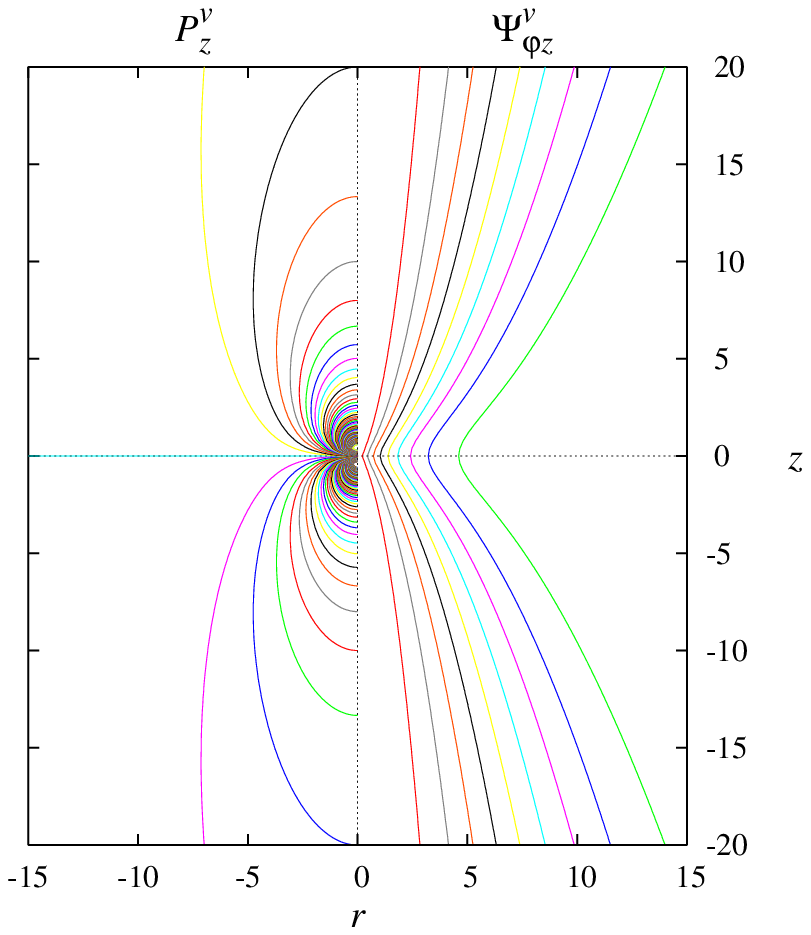}\put(-140,20){(\textit{a})}\includegraphics[bb=110bp 130bp 365bp 400bp,clip,width=0.33\columnwidth]{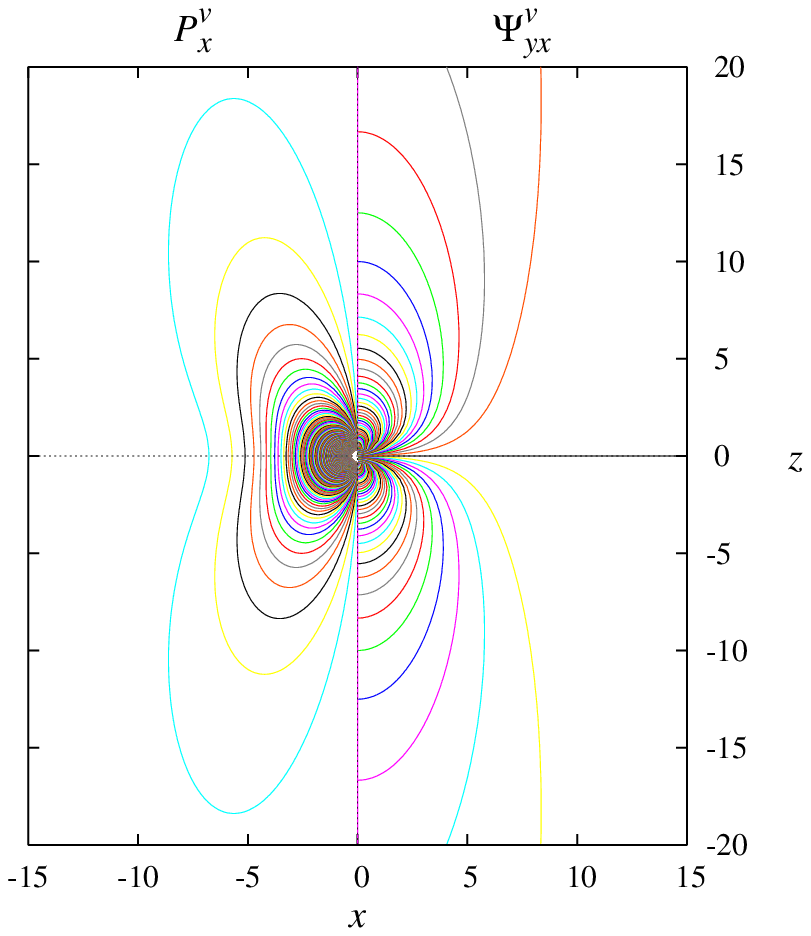}\put(-140,20){(\textit{b})}\includegraphics[bb=110bp 130bp 365bp 400bp,clip,width=0.33\columnwidth]{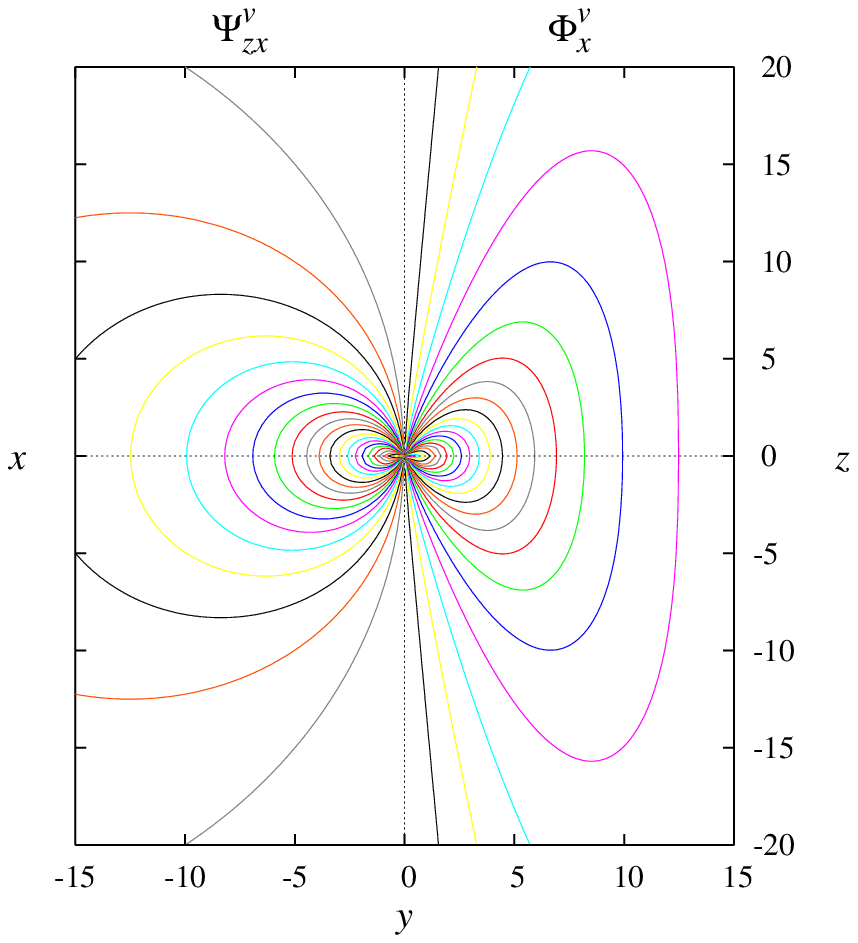}\put(-140,20){(\textit{c})}
\par\end{centering}

\caption{\label{fig:}Axial solutions for the pressure $P_{z}^{v}$ and the
stream function $\Psi_{z}^{v}$ shown at $r<0$ and $r>0,$ respectively
(\emph{a}). Pressure distribution $P_{x}^{v}$ and the transverse
stream function $\Psi_{yx}^{v}$ in the $xz$-plane shown at $x<0$
and $x>0,$ respectively, (\emph{b}), and axial stream function $\Psi_{yx}^{v}$
in the $xy$-plane $(y<0)$ and the electric potential $\Phi_{x}^{v}=-\Psi_{zx}^{v}$
$(y>0)$ in the $yz$-plane (\emph{c}).}
\end{figure}

Owing to the linearity of the problem, the mechanically induced flow
can be represented as a superposition of particular solutions due
to the force components parallel and perpendicular to the magnetic
field. In the following, these solutions are referred to as axial
and transverse ones. The former is obtained by taking $\vec{f}=\vec{e}_{z}$
in (\ref{eq:vpf}), which results in $\vec{\mathcal{L}}_{\{v,p,\phi\}}^{v}\vec{e}_{z}G=\{\vec{V},P,\Phi\}_{z}^{v},$
where 
\begin{equation}
\vec{V}_{z}^{v}=\vec{\nabla}\times(\vec{e}_{\varphi}\Psi_{\varphi z}^{v});\qquad P_{z}^{v}=-(3zR^{-1}\vec{\nabla}^{2}G+\partial_{z}G)/2;\qquad\Phi_{z}^{v}=0.\label{sol:V-axi}
\end{equation}
Axisymmetric poloidal velocity (\ref{sol:V-axi}) is defined by the
azimuthal component of the stream function $\Psi_{\varphi z}^{v}=-\partial_{r}G.$
The instantaneous streamlines of this flow, which are represented
by the isolines of $r\Psi_{\varphi z}^{v},$ are shown in Fig. \ref{fig:}(\emph{a})
together with the pressure distribution $P_{z}^{v}.$ Note that no
electric potential is induced because the e.m.f generated by such
an axisymmetric poloidal flow is purely azimuthal and, thus, solenoidal.
As seen, the flow is confined in the regions where the argument of
the exponential functions in (\ref{sol:d2G},\ref{sol:dzG}) is not
too small, i.e. $R\pm z\lesssim1.$ For the large axial distances
$|z|\gtrsim1$, these regions correspond to two parabolic wakes $r^{2}\lesssim|z|,$
which extend along the magnetic field in both directions from the
origin. In the mid-plane (|$z|\lesssim1),$ the solution drops off
exponentially over the cylindrical radius $r\sim1,$ which is the
dimensionless thickness of the Hartmann layer. Along the axis $(r=0)$,
(\ref{sol:V-axi}) yields
\begin{equation}
P_{z}^{v}=-(3\vec{\nabla}^{2}G+\partial_{z}G)/2=(2+\e^{-|z|})/8\pi z,\quad V_{zz}^{v}=r^{-1}\partial_{r}(r\partial_{r}G)=(1+z-\e^{-|z|})/8\pi z|z|,\label{sol:Pvz-z}
\end{equation}
according to which both the pressure and the axial velocity drop off
as $\sim z^{-1}.$ The incompressibility and the confinement of the
flow within the parabolic wakes of the radius $r\sim|z|^{1/2}$ results
in the radial velocity component decreasing as $V_{rz}^{v}\sim z^{-3/2}.$ 

\paragraph{Transverse force}

In this case solution is obtained by taking $\vec{f}=\vec{e}_{x},$
which results in $\vec{\mathcal{L}}_{\{v,p,\phi\}}^{v}\vec{e}_{x}G=\{\vec{V},P,\Phi\}_{x}^{v},$
where
\begin{equation}
\vec{V}_{x}^{v}=\vec{\nabla}\times(\vec{e}_{y}\Psi_{yx}^{v}+\vec{e}_{z}\Psi_{zx}^{v}),\quad P_{x}^{v}=\vec{e}_{x}\cdot\vec{\nabla}\vec{\nabla}^{2}G=-\frac{3x}{2R}\vec{\nabla}^{2}G,\quad\Phi_{x}^{v}=-\Psi_{zx}^{v}=\frac{y}{r}\partial_{r}G.\label{sol:V-trn}
\end{equation}
In this case, the flow, which is three-dimensional and has all three
velocity components, is described by two components of the stream
function. Isolines of the transverse component $\Psi_{yx}^{v}=\partial_{z}G$
represent the streamlines of the solenoidal part of the flow in the
planes parallel to both the magnetic field and the force, describes
the circulation in the $xz$-plane with the streamlines shown in Fig.
\ref{fig:}(\emph{b}) at $x>0.$ The $z$-component of the stream
function, which coincides with the induced electric potential in (\ref{sol:V-trn}),
describes circulation in the plane transverse to the magnetic field,
i.e. the $xy$-plane in the case under consideration. Streamlines
in this plane are shown in Fig. \ref{fig:}(\emph{c}) $(y<0)$ together
with the potential distribution in the $yz$-plane $(y>0).$ The pressure
in (\ref{sol:V-trn}), whose distribution in the $xz$-plane is plotted
in Fig. \ref{fig:}(\emph{b}) at $x<0,$ shows the characteristic
two-wake structure. Comparing this pressure with (\ref{sol:V-trn-z})
it is not hard to see that in the wakes with $r\sim z^{1/2}$ the
pressure drops off as $\sim z^{-3/2}.$ The transverse velocity component
along the axis is 
\begin{equation}
V_{xx}^{v}=\partial_{x}\Psi_{zx}^{v}-\partial_{z}\Psi_{yx}^{v}=-\vec{\nabla}^{2}G=(1+\e^{-|z|})/8\pi|z|,\label{sol:V-trn-z}
\end{equation}
which is in the same direction as the force. However, as seen in Fig.
(\ref{fig:})(\emph{b}) $(x>0)$, the circulation in the plane of
the magnetic field and force is directed oppositely to the latter
with the associated transverse velocity varying as 
\[
-\partial_{z}\Psi_{yx}^{v}=-\partial_{z}^{2}G=-(1-(1+|z|)\e^{-|z|})/8\pi z^{2},
\]
which at large distances drops off as $\sim z^{-2}.$ This decrease
is faster than $\sim z^{-1}$ for the total transverse velocity (\ref{sol:V-trn-z}),
which includes also the circulation transversely to the magnetic field
in the plane $xy$-plane. Thus, the latter obviously dominates at
large axial distances. 

The solutions for $\Phi_{x}^{v}$ and $\Psi_{zx}^{v},$ which according
to (\ref{sol:V-trn}) differ from each other only by the sign, are
substantially different from the solutions considered so far. The
difference is due to the first term in (\ref{sol:drG}), which as
noted above falls off at large cylindrical radii algebraically as
$\sim r^{-1}.$ In particular, for the mid-plane $z=0,$ the expressions
in (\ref{sol:V-trn}) reduce to
\[
\Phi_{x}^{v}=-\Psi_{zx}^{v}=\sin\varphi(\e^{-r/2}-1)/4\pi r,
\]
where $\sin\varphi=y/r.$ Thus, in contrast the other variables considered
above, $\Phi_{x}^{v}$ and $\Psi_{zx}^{v}$ fall off outside the wakes
as $\sim r^{-1}$ rather than exponentially, and thus, as seen in
Fig. \ref{fig:}(\emph{c}), they are not confined in the wakes.

\end{document}